\documentclass[draft,onecolumn,12pt]{IEEEtran}%journal

\usepackage{amsmath}   % From the American Mathematical Society

 \newtheorem{thm}{Theorem}%[subsection]
 \newtheorem{cor}{Corollary}
 \newtheorem{lem}{Lemma}
 \newtheorem{prop}{Proposition}
 \newtheorem{defn}{Definition}

 \newenvironment{prf}{{\emph{Proof: }}}
 {\hfill QED.}

\begin{document}
\title{New Lower Bounds on Sizes of Permutation Arrays}
\author{Lizhen Yang ,
        Kefei Chen, Luo Yuan
    % <-this % stops a space
\thanks{Manuscript received June 1, 2006.
        This work was supported by NFSE under grants 90104005 and 60573030.}% <-this % stops a space
\thanks{Lizhen Yang is with the department of computer science and
engineering, Shanghai Jiaotong University, 800 DongChuan Road,
Shanghai, R.P. China (fax: +86-21-34204221,
email:lizhen\_yang@msn.com).}
\thanks{Kefei Chen is with the department of computer science and
engineering, Shanghai Jiaotong University, 800 DongChuan Road,
Shanghai, R.P. China (fax: +86-21-34204221, email:
Chen-kf@sjtu.edu.cn).}}
% note the % following the last \IEEEmembership and also the first \thanks -
% instance, "A\textbf{} \textbf{}B" will typeset as "A B" not "AB". If you want
% "AB" then you have to do: "A\textbf{}\textbf{}B"
\markboth{Journal of \LaTeX\ Class Files,~Vol.~1, No.~11,~November~2002}{Shell \MakeLowercase{\textit{et al.}}: Bare Demo of IEEEtran.cls for Journals}
% *** Note that you probably will NOT want to include the author's name in ***
% *** the headers of peer review papers.                                   ***

% If you want to put a publisher's ID mark on the page
% (can leave text blank if you just want to see how the
% text height on the first page will be reduced by IEEE)
%\pubid{0000--0000/00\$00.00~\copyright~2002 IEEE}

%\specialpapernotice{(Invited Paper)}

\maketitle

\begin{abstract}
A permutation array(or code) of length $n$ and distance $d$,
denoted by $(n,d)$ PA, is a set of permutations $C$ from some
fixed set of $n$ elements such that the Hamming distance between
distinct members $\mathbf{x},\mathbf{y}\in C$ is at least $d$. Let
$P(n,d)$ denote the maximum size of an $(n,d)$ PA. This
correspondence focuses on the lower bound on $P(n,d)$. First we
give three improvements over the Gilbert-Varshamov lower bounds on
$P(n,d)$ by applying the graph theorem framework presented by
Jiang and Vardy. Next we show another two new improved bounds by
considering the covered balls intersections. Finally some new
lower bounds for certain values of $n$ and $d$ are given.
\end{abstract}

% Will be ignored for other modes.
\IEEEpeerreviewmaketitle

\section{Introduction}
\PARstart{L}{e}t $\Omega$ be an arbitrary nonempty infinite set.
Two distinct permutations $\mathbf{x},\mathbf{y}$ over $\Omega$
have distance $d$ if $\mathbf{x}\mathbf{y}^{-1}$ has exactly $d$
unfixed points.  A permutation array(permutation code, PA) of
length $n$ and distance $d$, denoted by $(n,d)$ PA, is a set of
permutations $C$ from some fixed set of $n$ elements such that the
distance between distinct members $\mathbf{x},\mathbf{y}\in C$ is
at least $d$. An $(n,d)$ PA of size $M$ is called an $(n,M,d)$ PA.
The maximum size of an $(n,d)$ PA is denoted as $P(n,d)$.

PAs are somewhat studies in the 1970s. A recent application by
Vinck ~\cite{Ferreira00,Vinck00Code,Vinck00Coded,Vinck00Coding} of
PAs to a coding/modulation scheme for communication over power
lines has created renewed interest in PAs. But there are still
many problems unsolved in PAs, e.g. one of the essential problem
is to compute the values of $P(n,d)$. It's known that determining
the exactly values of $P(n,d)$ is a difficult task, except for
special cases, it can be only to establish some lower bounds and
upper bounds on $P(n,d)$. We shall study how to determine lower
bound on $P(n,d)$ in this correspondence, and give some new
bounds.

\subsection{Concepts and Notations}
In this subsection, we introduce concepts and notations that will
be used throughout the correspondence.

Since for two sets $\Omega,\Omega'$ of the same size, the
symmetric groups $Sym(\Omega)$ and $Sym(\Omega')$ formed by the
permutations over $\Omega$ and $\Omega'$ respectively, under
compositions of mappings, are isomorphic, we need only to consider
the PAs over $Z_n=\{0,1,\ldots,n-1\}$ and write $S_n$ to denote
the special group $Sym(Z_n)$. In the rest of the correspondence,
without special pointed out, we always assume that PAs are over
$Z_n$. We also write a permutation $\mathbf{a}\in S_n$ as an
$n-$tuple $(a_0,a_1,\ldots,a_{n-1})$, where $a_i$ is the image of
$i$ under $\mathbf{a}$ for each $i$. Especially, we write the
identical permutation $(0,1,\ldots,n-1)$ as $\mathbf{1}$ for
convenience. The Hamming distance $d(\mathbf{a},\mathbf{b})$
between two $n-$tuples $\mathbf{a}$ and $\mathbf{b}$ is the number
of positions where they differ. Then the distance between any two
permutations $\mathbf{x},\mathbf{y}\in S_n$ is  equivalent to
their Hamming distance.

Let $C$ be an $(n,d)$ PA. A permutation in $C$ is also called a
codeword of $C$. For convenience for discussion, without loss of
generality, we always assume that $\mathbf{1}\in C$, and the
indies of an $n-$tuple (vector, array) are started by $0$. The
support of a binary vector
$\mathbf{a}=(a_0,a_1,\ldots,a_{n-1})\in\{0,1\}^n$ is defined as
the set $\{i:a_i=1,i\in Z_n\}$, and the weight of $\mathbf{a}$ is
the size of its support, namely the number of ones in
$\mathbf{a}$. The support of a permutation
$\mathbf{x}=(x_0,x_1,\ldots,x_{n-1})\in S_n$ is defined as the set
of the points not fixed by $\mathbf{x}$, namely $\{i\in Z_n:
x_i\neq i\}$=$\{i\in Z_n: \mathbf{x}(i)\neq i\}$, and the weight
of $\mathbf{x}$, denoted as $wt(\mathbf{x})$, is defined as the
size of its support, namely the number of points in $Z_n$ not
fixed by $\mathbf{x}$.

For an $(n,d)$ PA $C$, we say that a permutation $\mathbf{a}\in
S_n$ is covered by a codeword $\mathbf{x}\in C$, if
$d(\mathbf{a},\mathbf{x})<d$. The set of permutations in $S_n$
covered by $\mathbf{x}\in C$ is denoted as
$\mathbf{B}(\mathbf{x})$ and called the covered ball of
$\mathbf{x}$. A derangement of order $k$ is an element of $S_k$
with no fixed points. Let $D_k$ be the number of derangements of
order $k$, with the convention that $D_0=1$. Then
$D_k=k!\sum_{i=0}^k\frac{(-1)^k}{k!}=\left[\frac{k!}{e}\right]$,
where $[x]$ is the nearest integer function, and  $e$ is the base
of the natural logarithm. Then
\begin{equation}\label{eq:vol-PA}
|\mathbf{B}(\mathbf{x})|=V(n,d-1)=\sum_{i=0}^{d-1}{n\choose i}D_i.
\end{equation}
For an arbitrary permutation $\mathbf{x}\in S_n$,
$d(\mathbf{x},C)$ stands for the Hamming distance between
$\mathbf{x}$ and $C$, i.e., $d(\mathbf{x},C)=\min_{\mathbf{c}\in
C}d(\mathbf{x},\mathbf{c})$. A permutation $\mathbf{x}$ is called
covered by $C$ if $d(\mathbf{x},C)<d$. The set of permutations
covered by $C$ is denoted as $\mathbf{B}(C)$ and called the
covered ball of $C$. Clearly, $\mathbf{B}(C)=\cup_{\mathbf{c}\in
C}\mathbf{B}(\mathbf{c})$.

Finally, we define $P[n,d-1]$ as the maximum size of the subset
$\Gamma$ of $S_n$ such that the distance between two distinct
permutations in $\Gamma$ is $d-1$ at most. We will show that
$P(n,d)$ have close relations with $P[n,d-1]$.

\subsection{Previous Work on the Lower Bounds on $P(n,d)$}
By the definitions of $P(n,d)$, it is easy to obtain the following
well-known elementary consequences that are firstly appeared in
~\cite{Deza78} and summarized in ~\cite{wensong04}.
\begin{prop}\label{prop:elementay:conse:P(n,d)}
\begin{eqnarray}
P(n,2)&=&n!,\\
P(n,3)&=&n!/2,\\
P(n,n)&=&n,\\
P(n,d)&\geq& P(n-1,d),P(n,d+1),\label{eq:n,d,d+1}\\
P(n,d)&\leq& nP(n-1,d),\label{eq:n,n-1:prop:elementay:conse:P(n,d)}\\
P(n,d)&\leq& n!/(d-1)!.
\end{eqnarray}
\end{prop}

A latin square of order $n$ is an $(n,n)$ PA. Two latin squares
$L=(L_{i,j})$ and $L'=(L'_{i,j})$ are orthogonal if
$\{(L_{i,j},L'_{i,j}):1\leq i,j\leq n\}=\{1,2,\ldots,n\}^2$. The
following proposition was proved by Colbourn et
al.~\cite{Colbourn04}.

\begin{prop}~\cite{Colbourn04}.
If there are $m$ mutually orthogonal latin squares of order $n$,
then $P(n,n-1)\geq mn$. In particular, if $q$ is a prime-power,
then $P(q,q-1)=q(q-1)$.
\end{prop}

It was pointed out by Frankl and Deza~\cite{Frankl77} that the
existence of a sharply $k-$transitive group acting on a set of
size $n$ is equivalent to a maximum $(n,n-k+1)$ PA. It is well
known that the group $PGL(2,q)$, consisting of fractional linear
transformations $x\mapsto (ax+b)/(cx+d),ad-bc\neq 0$, is sharply
$3-$transitive acting on $X=F_q\cup \{\infty\}$, and the Mathieu
groups $M_{11}$ and $M_{12}$ are sharply $4-$ and $5-$transitive
on sets of size 11 and 12, respectively.

\begin{prop}~\cite{Frankl77}.\label{prop:PA_sharply}
If $q$ is a prime-power, then $P(q+1,q-1)=(q+1)q(q-1)$.
Additionally, $P(11,8)=11\cdot 10\cdot 9\cdot 8$ and
$P(12,8)=12\cdot 11\cdot 10\cdot 9\cdot 8$.
\end{prop}

Let $F_q$ be a finite field of order $q$. A polynomial $f$ over
$F_q$ is a permutation polynomial if the mapping it defines is
one-to-one. Let $N_d(q)$ denote the number of the permutation
polynomials over $F_q$ of given degree $d\geq 1$. By a direct
construction of PAs from permutation polynomials, Chu et
al.~\cite{wensong04} proved the following connection between
$P(q,q-d)$ and $N_i(q)$.

\begin{prop}~\cite{wensong04}.\label{prop:wensong:permutation-poly}
Let $q$ be a prime power. Then $P(q,q-d)\geq \sum_{i=1}^d N_i(q)$.
\end{prop}

Unfortunately, not much is known about permutation polynomials.
While their classification and enumeration are far from complete,
everything is known for $d<6$. The normalized permutation
polynomials with degree $d\leq 5$, together with the total
produced by each class are given in Table
~\ref{table:norm-per-poly}, summarized by Chu et
al.~\cite{wensong04} according to the table in
~\cite{Lidl-finite-97}.

\begin{table}
  \centering
  \begin{tabular}{|l|l|l|}
    \hline
    Normalized Permutation Polynomials & $q$ restriction & Total \\
    \hline
    $x$ & any $q$ & $q(q-1)$ \\
    $x^2$& $q\equiv 0\mod 2$&$q(q-1)$\\
    $x^3$&$q\not\equiv 1\mod 3$&$q^2(q-1)$ or $q(q-1)$\\
    $x^3-ax$($a$ not a square)&$q\equiv 0\mod 3$&$q(q-1)^2/2$\\
    $x^4\pm3x$&$q=7$&$2q^2(q-1)$\\
    $x^4+a_1x^2+a_2x$(if only root in $F_q$ is 0)&$q\equiv 0\mod 2$&$\frac{1}{3}q(q-1)(q^2+2)$\\
    $x^5$&$q\not\equiv 1\mod 5$&$q^2(q-1)$ or $q(q-1)$\\
    $x^5-ax$($a$ not a fourth power)&$q\equiv 0\mod 5$&$\frac{3}{4}q(q-1)^2$\\
    $x^5+ax(a^2=2)$&$q=9$&$2q^2(q-1)$\\
    $x^5\pm 2x^2$&$q=7$&$2q^2(q-1)$\\
    $x^5+ax^3\pm x^2+3a^2x$($a$ not a square)&$q=7$&$q^2(q-1)^2$\\
    $x^5+ax^3+5^{-1}a^2x$($a$ arbitrary)&$q\equiv\pm 2\mod 5$&$q^3(q-1)$\\
    $x^5+ax^3+3a^2x$($a$ not a square)&$q=13$&$\frac{1}{2}q^2(q-1)^2$\\
    $x^5-2ax^3+a^2x$($a$ not a square)&$q\equiv 0\mod 5$&$\frac{1}{2}q^2(q-1)^2$\\
    \hline
  \end{tabular}
  \caption{Normalized Permutation Polynomials with degree $d\leq 5$}\label{table:norm-per-poly}
\end{table}

By a simply observation,  Chu et al.~\cite{wensong04} also proved
another connection between permutation polynomials and $P(q,q-d)$.

\begin{prop}~\cite{wensong04}.\label{prop:wensong04:PP:d+1}
Suppose $q$ is a prime-power and that there are $M$ monic
permutation polynomial over $F_q$ of degree less than or equal to
$d+1$. Then $P(q,q-d)\geq M$.
\end{prop}

The following result is immediately gotten from
Proposition~\ref{prop:wensong04:PP:d+1} and
Table~\ref{table:norm-per-poly}.
\begin{cor}~\cite{wensong04}.\label{cor:wensong04:q-2}
If $q$ is a prime-power, $q\not\equiv 1\mod 3$~\footnote{In
~\cite{wensong04}, $q\not\equiv 1\mod 3$ is replaced by
$q\not\equiv 2\mod 3$, but by Table~\ref{table:norm-per-poly} it
should be $q\not\equiv 1\mod 3$.}, then $P(q,q-2)\geq q^2$.
\end{cor}

In ~\cite{Klove00combin}, T.Kl{\o}ve proved the following lower
bound on $P(n,n-1)$ by generalized the approach in ~\cite{Deza78}.
\begin{prop}\label{prop:n,n-1}
Let $n=\sum_{i=1}^u p_i^{c_i}$ be the standard factorization of
$n$, and let
\begin{equation}\label{eq:theta_function}
\theta(n)=\min\{p_i^{c_i}|1\leq i\leq u\}.
\end{equation}
Then for all $n>1$ we have
\[
P(n,n-1)\geq n(\theta(n)-1).
\]
\end{prop}

The other explicit constructions leading to lower bounds on
$P(n,d)$ are listed below. In~\cite{Ding20}, C. Ding, et al.
presented a construction of $(mn, mn-1)$ PA with size $m|C|$ from
an $r-$bounded $(n,n-1)$ PA and an $s-$separable $(m,m-1)$ PA.
In~\cite{Fu-fang-wei04}, Fu and Kl{\o}ve presented two
constructions of PAs with length $qn$ from $(n,d;q)$ codes and
$(n,d)$ PAs. In~\cite{wensong04}, Chu et al. proposed a recursive
construction of PA and used this construction to derive a lower
bound on $P(n,4)$ and a lower bound that $P(n,n-2)\geq 2q(q-1)$,
whenever $n=q+q'$ is a sum of two prime powers with $0\leq
q'-q\leq
 2$.

For certain small values of $n$ and $d$, the lower bounds on
$P(n,d)$ can be also directly determined by computational
constructions. Deza and Vanstone~\cite{Deza78} first used computer
construction to prove $P(6,5)=18$ and $P(10,9)\geq 32$. In
~\cite{wensong04}, Chu et al. presented three computational
methods of  clique search, greedy algorithm and automorphisms, and
got some new lower bounds for certain values of $n$ and $d$.

For $n\leq 13$ and certain values of $n\geq 14$ and $d$, the best
previous lower bounds on $P(n,d)$ are summarized in
~\cite{wensong04}.

The only general lower bound on $P(n,d)$ is the Gilbert-Varshmov
bound, which is derived in a similar way as the Gilbert-Varshmov
bound for binary codes. Let $A(n,d)$ be the maximum size of an
$(n,d)$ binary code, then $$A(n,d)\geq \frac{2^n}{V_2(n,d-1)},$$
where $V_2(n,d-1)$ is the volume of a sphere in $\{0,1\}^n$ of
radius $d-1$, that is,
\begin{equation}\label{eq:vol:binarycode}
V_2(n,d-1)=\sum_{i=0}^{d-1}{n\choose i}.
\end{equation}
Similarly, the Gilbert-Varshamov bound~\cite{Frankl77} on $P(n,d)$
is as follows:
\[P(n,d)\geq \frac{n!}{V(n,d-1)}.\]

\subsection{Our New Results}
In this correspondence, we first give three improvements over the
Gilbert-Varshamov lower bounds on $P(n,d)$ by using the graph
theorem framework presented by Jiang and Vardy in
~\cite{TaoJiang04}. In 2004, Jiang and Vardy presented a graph
theorem framework which may lead to improvements over
Gilbert-Varshamov bound for codes if the corresponding
Gilbert-Vashamov graphs are sparse. They were successful to
asymptotically improve the Gilbert-varshamov bound on size of
binary codes by a factor of $n$ when $d$ is proportional to $n$,
namely, $d=\alpha n$ for some positive constant $\alpha$.
Recently, Vu and Wu~\cite{Van05} generalized the results of Jiang
and Vardy to $q$-ary codes. Employing the graph theorem framework,
we also establish the following three new theorems in lower bounds
on $P(n,d)$.
\begin{thm}\label{thm:graph:presice}
For $x\in R$, let $\lceil x\rceil^{+}$ denote the smallest
nonnegative integer $m$ with $m\geq x$. Given positive integers
$n$ and $d$, with $d\leq n$, let $E(n,d)$ denote the following
quantity:
\[
E(n,d-1)=\frac{1}{6}\sum_{i=2}^{d-1}\sum_{j=2}^{d-1} {n\choose
i}D_iL_{i,j}
\]
where
\[
L_{i,j}=\sum_{k=\lceil\frac{i+j-d+1}{2}\rceil^{+}}^{\min(i,j)}\sum_{l=0}^{\min\{d+2k-i-j-1,k\}}{i\choose
k}{n-i\choose j-k}{k\choose l}(l+j-k)!.
\]
Then
\begin{equation}\label{eq:graph:presice}
P(n,d)\geq \frac{n!}{10V(n,d-1)}\left(\log_2 V(n,d-1)-1/2\log_2
E(n,d-1) \right)
\end{equation}
\end{thm}

\begin{thm}\label{thm:graph:d/n}
Let $\alpha$ be a constant satisfying $0<\alpha<1/2$. Then there
is a positive constant $c$ depending on $\alpha$ such that the
following holds. For $d=\alpha n$,
\[
P(n,d)\geq c\frac{n!}{V(n,d-1)}\log_2V_2(n,d-1).
\]
\end{thm}

\begin{thm}\label{thm:graph:d:small}
Let $\alpha$ be a constant satisfying $0< \alpha< 1$. Then there
is a positive constant $c$ depending on $\alpha$ such that the
following holds. For $d= n^{\alpha}$,
\[
P(n,d)\geq c\frac{n!}{V(n,d-1)}\log_2V(n,d-1).
\]
\end{thm}

Secondly, another two improvements over Gilbert-Varshamov lower
bounds are established by considering the covered balls
intersections. We will prove in section III that
\[
P(n,d)\geq \frac{2\cdot n!}{V(n,d-1)+P[n,d-1]}.
\]
Let $C'$ be an $(n,M,d)$ PA, then we will prove in section III
that
\[
P(n,d)\geq \frac{n!M}{|\mathbf{B}(C')|}.
\]

Our third contribution is to give some new lower bounds on
$P(n,d)$ for certain cases of $n$ and $d$ based on the two new
relations:

for $n\geq d> 3$
\begin{equation}
P(n-1,n-3)\geq P(n,d),
\end{equation}
and for $n\geq d> 2$
\[
P(n-1,d-2)\geq \frac{2}{n}P(n,d).
\]

\section{Improved Gilbert-Varshamov Bound by Graph Theoretic Framework}
We first recall a few basic notions from graph theory. A graph $G$
consists of a (finite) set $V(G)$ of vertices and a set $E(G)$ of
edges, where an edge is a (non-ordered) pair $(a, b)$ with $a,
b\in V(G)$. If $a$ and $b$ form an edge, we say that they are
adjacent. The set of all neighbors of a vertex $v$ is denoted as
$N(v)$ and called the neighborhood of $v$. The  degree of a vertex
$v\in V(G)$, denoted as $\deg(v)$, is defined as $\deg(v)=|N(v)|$.
The graph is $D$-regular if the degree of every vertex equals $D$.
A subset $I$ of $V(G)$ is an independent set if it does not
contain any edge. The independence number of $G$ is the size of
the largest independent set in $G$, and is denoted as $\alpha(G)$.

\begin{defn}Let $n$ and $d\leq n$ be positive integers. The corresponding
Gilbert graph $\mathcal{G}_2$ over $\{0,1\}^n$ is defined as
following: $V(\mathcal{G}_2)=\{0,1\}^n$ and
$\{\mathbf{u},\mathbf{v}\}\in E(\mathcal{G}_2)$ if and only if
$1\leq d(\mathbf{u},\mathbf{v})\leq d-1$.
\end{defn}

\begin{defn}Let $n$ and $d\leq n$ be positive integers. The corresponding
Gilbert graph $\mathcal{G}_P$  over $S_n$  is defined as
following: $V(\mathcal{G}_P)=S_n$ and
$\{\mathbf{u},\mathbf{v}\}\in E(\mathcal{G}_P)$  if and only if
$1\leq d(\mathbf{u},\mathbf{v})\leq d-1$.
\end{defn}

Then clearly, an $(n,d)$ binary code is an independent set in the
Gilbert graph $\mathcal{G}_2$. Conversely, any independent set in
$\mathcal{G}_2$ is an $(n,d)$ binary code. This means
$A(n,d)=\alpha(\mathcal{G}_2)$. Similarly,
$P(n,d)=\alpha(\mathcal{G}_P)$. By applying a simple observation
on graph to a graph theorem in Bollob\'{a}s~\cite[Lemma 15,
p.296]{Bolloas85}, Jiang and Vardy~\cite{TaoJiang04} prove the
following theorem.
\begin{thm}~\cite{TaoJiang04}.\label{thm:general-graph}
Let $G$ be a graph with maximum degree at most $D$, and suppose
that for all $\mathbf{v}\in V(G)$, the subgraph of $G$ induced by
the neighborhood of $\mathbf{v}$ has at most $T$ edges. Then
\[
\alpha(G)\geq \frac{n(G)}{10D}\left(\log_2 D
-1/2\log_2(T/3)\right),
\]
where $n(G)$ is the number of vertices of $G$.
\end{thm}

We consider the Hamming sphere graph $\mathcal{G}_{SP}$ over $S_n$
that is the subgraph of the Gilbert graph $\mathcal{G}_P$ over
$S_n$ induced by the neighborhood $N(\mathbf{1})$ of the vertex
$\mathbf{1}\in V(\mathcal{G}_P)$. Clearly, the subgraph induced in
the Gilbert graph over $S_n$ by the neighborhood of any other
vertex in $\mathcal{G}_P$ is isomorphic to $\mathcal{G}_{SP}$. To
derive an upper bound for the edges of $\mathcal{G}_{SP}$, we need
to consider the Hamming sphere graph $\mathbf{G}_{S2}$ over
$\{0,1\}^n$, that is the subgraph of the Gilbert graph
$\mathcal{G}_2$ over $\{0,1\}^n$ induced by the neighborhood
$N(\mathbf{0})$ of the vertex $\mathbf{0}\in V(\mathcal{G}_2)$.
For the sake of clearer presentation, we define
$T=|E(\mathcal{G}_{SP})|,
 D=|V(G_{SP})|=V(n,d-1)-1, T'=|E(\mathcal{G}_{S2})|,
D'=|V(G_{S2})|=V_2(n,d-1)-1$, where $V(n,d-1)$ and $V_2(n,d-1)$
are defined by (\ref{eq:vol-PA}) and (\ref{eq:vol:binarycode})
respectively.

\begin{lem}\label{lem:bound:L_i,j}
For any $\mathbf{x}\in S_n$ of weight $i$, there are at most
\[
L_{i,j}=\sum_{k=\lceil\frac{i+j-d+1}{2}\rceil^{+}}^{\min(i,j)}\sum_{l=0}^{\min\{d+2k-i-j-1,k\}}{i\choose
k}{n-i\choose j-k}{k\choose l}(l+j-k)!
\]
permutations of weight $j$ with distance less than $d$ to
$\mathbf{x}$, where $\lceil x\rceil^{+}$ denotes the smallest
nonnegative integer not less than $x$.
\end{lem}
\begin{prf}
Without loss of generality, suppose the support of $\mathbf{x}$ is
$X=\{0,1,\ldots,i-1\}$. Let $\mathbf{y}$ be an arbitrary
permutation with weight of $j$ and support of $Y$, having distance
less than $d$ to $\mathbf{x}$. Let $Z=X\cap Y$ and $R=\{r\in X\cap
Y: \mathbf{x}(r)\neq \mathbf{y}(r)\}$.  Then it follows from
$d-1\geq d(\mathbf{x},\mathbf{y})=|X|+|Y|-2|X\cap
Y|+|R|=i+j-2|Z|+|R|$ that
\begin{equation}\label{eq:lem:bound:L_i,j}
|R|\leq d+2|Z|-i-j-1.
\end{equation}
Since $R\geq 0$, $|Z|\geq \frac{i+j-d+1}{2}$ by
(\ref{eq:lem:bound:L_i,j}). There are at most ${i\choose
k}{n-i\choose j-k}$ candidates of $Y$ such that $|Z|=k$, and for
each candidate of $Y$ satisfying $|Z|=k$ there are at most
${k\choose l}(l+j-k)!$ corresponding permutations satisfying
$|R|=l$. Therefore the lemma follows immediately.
\end{prf}

\begin{lem}\label{lem:bound:T}
\begin{equation}\label{eq:lem:bound:T}
T\leq \frac{1}{2}\sum_{i=2}^{d-1}\sum_{j=2}^{d-1} {n\choose
i}D_iL_{i,j}
\end{equation}
\end{lem}
\begin{prf}
Since $\mathcal{G}_{SP}$ has ${n\choose i}D_i$ vertices of weight
$i$, and there has no vertices with weight $1$,
(\ref{eq:lem:bound:T}) follows immediately from
Lemma~\ref{lem:bound:L_i,j}.
\end{prf}

Comparing the foregoing expression for the upper bound on $T$ with
the expression for $E(n,d-1)$ in Theorem~\ref{thm:graph:presice},
we see that $E(n,d-1)\geq \frac{T}{3}$. Thus
Lemma~\ref{lem:bound:T} in conjunction with
Theorem~\ref{thm:general-graph} induces (\ref{eq:graph:presice}).
This completes the proof of Theorem~\ref{thm:graph:presice}.

Now we turn to the asymptotic bounds on $T$ which will in turn
induce the asymptotic bounds on $P(n,d)$. Instead of using the
 upper bound on $T$ presented in
Lemma~\ref{lem:bound:T}, we use the following upper bound on $T$
which is more weaker but more  easily to be treated.
%%%%%%%%%%%%%%%%%%%%%%%%%%%%%%%%%%%%%%%%%%%%%%%%%%%%%%%%%%%%%%%%%%%%
\begin{lem}\label{lem:relation:T:T' and D'}
\[
T\leq (T'+D')D_{d-1}^2.
\]
\end{lem}
\begin{prf}
Let $\mathbf{x}$ and $\mathbf{y}$ be an arbitrary pair of adjacent
vertices in $\mathcal{G}_{SP}$ with supports $X$ and $Y$
respectively. Then $d(\mathbf{x},\mathbf{y})\leq d-1$. Since they
take differ values in points of $(X\cup Y)/(X\cap Y)$,
$d(\mathbf{x},\mathbf{y})\geq |(X\cup Y)/(X\cap
Y)|=|X|+|Y|-2|X\cap Y|$. Clearly, an binary vector
 is uniquely determined by its support. Let $\mathbf{x}',
\mathbf{y}'\in \{0,1\}^n$ with supports $X,Y$ respectively. Then
\[d(\mathbf{x}',\mathbf{y}')=|X|+|Y|-2|X\cap
Y|\leq d(\mathbf{x},\mathbf{y})\leq d-1.\]
Furthermore,
\[
d(\mathbf{x}',\mathbf{0})=|X|=wt(\mathbf{x})=d(\mathbf{x},\mathbf{1})\leq
d-1,
\]
thereby $\mathbf{x}'\in \mathcal{G}_2$, similarly, $\mathbf{y}'\in
\mathcal{G}_2$. Hence $(\mathbf{x}',\mathbf{y}')\in
E(\mathcal{G}_2)$ whenever $X\neq Y$. Therefore
\begin{eqnarray*}
|\{(X,Y): X,Y \mbox{ are supports of a pair of adjacent vertices
in }\mathcal{G}_{SP} \mbox{ with } X\neq Y\}|&\leq&
|E(\mathcal{G}_2)|\\&=&T',
\end{eqnarray*}
\begin{eqnarray*}
|\{(X,Y): X,Y \mbox{ are supports of a pair of adjacent vertices
in }\mathcal{G}_{SP} \mbox{ with } X=Y\}|&\leq&
|V(\mathcal{G}_2)|\\&=&D'.
\end{eqnarray*}
Then
\[
|\{(X,Y): X,Y \mbox{ are supports of a pair of
adjacent vertices in }\mathcal{G}_{SP}\}|\leq T'+D',
\]
which in conjunction with the fact
$$|\{\mathbf{x}:\mathbf{x}\in \mathcal{G}_{SP} \mbox { with
support } X \}|\leq D_{d-1}$$ completes the proof.
\end{prf}

Vu and Wu~\cite{Van05} have proved the following relation between
$T'$ and $D'$.
\begin{lem}\label{lem:relation_binary_Gilbert}
For every constant $0<\alpha<1/2$ there is a positive constant
$\epsilon$ such that the following holds: for $d=\alpha n$,
\[
T'\leq D'^{2-\epsilon}.
\]
\end{lem}

\begin{lem}\label{lem:compare:poly:D}
For any positive constant $\epsilon$, $0<\alpha<1$ and any
polynomial function $f(x)$, there exists a positive value $N$, for
$n\geq N$, $f(n)\leq {n\choose d-1}^{\epsilon}$, whenever
$d=\alpha n$.
\end{lem}
\begin{prf}
It is well known that
  $$\lim_{n\to\infty}\frac{{n\choose
d}}{\frac{1}{\sqrt{2n\pi
\alpha(1-\alpha)}}\left(\frac{1}{\alpha^{\alpha}(1-\alpha)^{1-\alpha}}\right)^n}=1,$$
then
\begin{eqnarray*}
\lim_{n\to \infty}\frac{f(n)}{{n\choose
d-1}^{\epsilon}}&=&\lim_{n\to \infty}\frac{f(n)}{{n\choose
d}^{\epsilon}}\cdot \frac{{n\choose d}^{\epsilon}}{{n\choose d-1}^{\epsilon}}\\
&=&\lim_{n\to \infty}f(n)\left(\sqrt{2n\pi
\alpha(1-\alpha)}(\alpha^{\alpha}(1-\alpha)^{1-\alpha})^n\right)^{\epsilon}\left(\frac{n-d+1}{d}\right)^{\epsilon}\\
&=&0,
\end{eqnarray*}
which implies the statement.
\end{prf}

\begin{lem}\label{lem:relation_permutation_Gilbert_1}
For every constant $0<\alpha<1/2$ there is a positive constant
$\epsilon$ such that the following holds: for $d=\alpha n$,
\[
T\leq \frac{D^2}{D'^{\epsilon}}.
\]
\end{lem}
\begin{prf}It follows from Lemma~\ref{lem:relation:T:T' and D'}  that $T\leq
(T'+D')D_{d-1}^2$, and while it follows from  the definitions of
$D$ and $D'$ that $D\geq {n\choose d-1}D_{d-1}> D'D_{d-1}/d$. So
we have
\begin{equation*}
\begin{array}{lcl}
\frac{D^2}{T}&\geq&\frac{\left(D'D_{d-1}/d\right)^2}{(T'+D')D_{d-1}^2}\\
&=&\frac{D'^2}{d^2(T'+D')}.
\end{array}
\end{equation*}
Then by Lemma~\ref{lem:relation_binary_Gilbert} there exists a
positive constant $\epsilon$ such that
\begin{equation}\label{eq:prf:lem:relation_permutation_Gilbert_1:1}
\frac{D^2}{T}\geq
\frac{D'^2}{d^2(D'^{2-\epsilon}+D')}=\frac{D'^{\epsilon}}{d^2(1+D'^{\epsilon-1})}\geq
\frac{D'^{\varepsilon}}{2d^2}
\end{equation}
where $\varepsilon=\min(\epsilon,1)$.  By
Lemma~\ref{lem:compare:poly:D}, there exists a positive constant
$N$ such that for $n\geq N$, $2d^2=2\alpha^2 n^2<({n\choose
d-1})^{\varepsilon/2}<D'^{\varepsilon/2}$, which in conjunction
with (\ref{eq:prf:lem:relation_permutation_Gilbert_1:1}) implies
$\frac{D^2}{T}\geq D'^{\varepsilon/2}$. Since $\frac{D^2}{T}>1$
always holds, there exists a positive constant $\varepsilon'$ such
that for $0<n<N$, $\frac{D^2}{T}>D'^{\varepsilon'}$. Taking
$\epsilon'=\min(\varepsilon/2,\varepsilon')$, then for all
$d=\alpha n$, $\frac{D^2}{T}\geq D'^{\epsilon'}$, namely $T\leq
\frac{D^2}{D'^{\epsilon'}}$.
\end{prf}

\emph{Proof of Theorem\ref{thm:graph:d/n}:} We are now ready to
complete the proof of Theorem~\ref{thm:graph:d/n}. Let $\alpha$ be
a constant satisfying  $0<\alpha <1/2$. Then by the definitions of
$D$ and $T$, Theorem~\ref{thm:general-graph} and
Lemma~\ref{lem:relation_permutation_Gilbert_1},  for case
$d=\alpha n$ there exists a positive constant $\epsilon$ such that
\begin{equation*}
\begin{array}{lcl}
\alpha(G_{P})&\geq&
\frac{n!}{10D}\left(\log_2D-1/2\log_2\left(\frac{D^{2}}{3D'^{\epsilon}}\right)\right)\\
&\geq&\frac{\min(\epsilon,1)}{20}\cdot\frac{n!}{D}(\log_2 D'+\log_2 3)\\
&\geq&\frac{\min(\epsilon,1)}{20}\cdot \frac{n!}{V(n,d-1)}\log_2
V_2(n,d-1).
\end{array}
\end{equation*}
Then we complete the proof. \hfill{QED.}

%%%%%%%%%%for case d=n^{\alpha}

\begin{lem}\label{lem:relation_permutation_Gilbert_2}
For every constant $0< \alpha <1$ there is a positive constant
$\epsilon$ such that whenever $d=n^{\alpha}$,
\[
T\leq D^{2-\epsilon}.
\]
\end{lem}
\begin{prf}The proof relies on the following three lemmas.

\begin{lem}\label{lem:relation_binary_Gilbert:d=n^a}
For every constant $0<\alpha<1$ there is a positive constant
$\epsilon$ such that the following holds: for $d=n^{\alpha}$,
\[
T'\leq D'^{2-\epsilon}.
\]
\end{lem}
\begin{prf}
Let $\alpha'$ be a constant satisfying $0<\alpha'<1/2$. Suppose
the Hamming sphere graphs over $\{0,1\}^n$ defined for
$d=n^{\alpha}$  and $d=\alpha'n$ are $\mathcal{G}'_{S_2}$ and
$\mathcal{G}''_{S_2}$ respectively. Let
$T'=|E(\mathcal{G}'_{S_2})|$, $T''=|E(\mathcal{G}''_{S_2})|$ and
$D'=|V(\mathcal{G}'_{S_2})|=|V(\mathcal{G}''_{S_2})|=V_2(n,d-1)-1$.
Clearly, there exists a positive integer $N$ such that for $n\geq
N$, $n^{\alpha}\leq \alpha' n$. This implies that for $n\geq N$,
$E(\mathcal{G}'_{S_2})\subseteq E(\mathcal{G}''_{S_2})$, which
means $T'\leq T''$. Then by
lemma~\ref{lem:relation_binary_Gilbert}, there exists a positive
constant $\epsilon'$ such that
\[
T'\leq T''\leq D'^{2-\epsilon'},
\]
whenever $n\geq N$. Moreover, $T'<D'^2$ always holds, then there
exists a positive constant $\epsilon''$ such that
\[
T'\leq D'^{2-\epsilon''}
\]
for $0<n<N$. Taking $\epsilon=\min\{\epsilon',\epsilon''\}$, then
$T'\leq D'^{2-\epsilon}$.
\end{prf}

\begin{lem}\label{lem:forprf:lem:relation_permutation_Gilbert_2}
For every pair of constants $0< \alpha <1$ and $0< \delta<1$
satisfying $1-\delta-\alpha>0$, whenever $d=n^{\alpha}$,
\[
\lim_{n\to \infty}\frac{D_{d-1}}{ D^{1-\delta}}=0.
\]
\end{lem}
\begin{prf}By the definitions of $D$ and $D_{d-1}$ we have
\begin{equation}
\begin{array}{lcl}
\lim\limits_{n\to
\infty}\frac{D_{d-1}}{D^{1-\delta}}&\leq&\lim\limits_{n\to
\infty}\frac{D_{d-1}}{\left({n\choose
d-1}D_{d-1}\right)^{1-\delta}}\\
&=&\lim\limits_{n\to
\infty}\frac{D_{d-1}^{\delta}}{\left({n\choose
d-1}\right)^{1-\delta}}\\
&=&\lim\limits_{n\to
\infty}\frac{((d-1)!/e)^{\delta}}{\left(\frac{n!}{(d-1)!(n-d+1)}\right)^{1-\delta}}\\
&=&\lim\limits_{n\to
\infty}c\frac{(d-1)!(n-d+1)^{1-\delta}}{n!^{1-\delta}}
\end{array}
\end{equation}
where constant $c=e^{-\delta}$. Then from Stirling's formula
$\lim_{n\to \infty}\frac{n!}{\sqrt{2\pi
n}\left(\frac{n}{e}\right)^n}=1$ it follows
\begin{eqnarray}
%\begin{array}{lcl}
\lim\limits_{n\to \infty}\frac{D_{d-1}}{D^{1-\delta}}&\leq&
\lim\limits_{n\to
\infty}c\frac{\sqrt{2\pi(d-1)}\left(\frac{d-1}{e}\right)^{d-1}\left(\sqrt{2\pi (n-d+1)}\left(\frac{n-d+1}{e}\right)^{n-d+1}\right)^{1-\delta}}{\left(\sqrt{2\pi n}\left(\frac{n}{e}\right)^{n}\right)^{1-\delta}}\nonumber\\
&=&\lim\limits_{n\to \infty}c\sqrt{2\pi
(d-1)}\left(\frac{n-d+1}{n}\right)^{\frac{1-\delta}{2}}\frac{\left(\frac{d-1}{e}\right)^{d-1}\left(\frac{n-d+1}{e}\right)^{(n-d+1)(1-\delta)}}
{\left(\frac{n}{e}\right)^{n(1-\delta)}}\nonumber\\
&\leq&\lim\limits_{n\to \infty}c\sqrt{2\pi
(d-1)}\frac{\left(\frac{d-1}{e}\right)^{d-1}\left(\frac{n-d+1}{e}\right)^{(n-d+1)(1-\delta)}}
{\left(\frac{n}{e}\right)^{n(1-\delta)}}\label{eq:lem:forprf:lem:relation_permutation_Gilbert_22}
\end{eqnarray}
By multiplying
$$\lim_{n\to\infty}\frac{e^{-1}n^{\alpha(d-1)}}{(d-1)^{d-1}}=e^{-1}\lim_{n\to\infty}\left(1+\frac{1}{d-1}\right)^{d-1}=e^{-1}e=1$$
and inequality $n-d+1\leq n$,
(\ref{eq:lem:forprf:lem:relation_permutation_Gilbert_22}) yields
\begin{eqnarray*}
\lim\limits_{n\to
\infty}\frac{D_{d-1}}{D^{1-\delta}}&\leq&\lim\limits_{n\to
\infty}ce^{-1}\sqrt{2\pi (d-1)}
\frac{\left(\frac{n^{\alpha}}{e}\right)^{d-1}\left(\frac{n}{e}\right)^{(n-d+1)(1-\delta)}}{\left(\frac{n}{e}\right)^{n(1-\delta)}}\\
&=&\lim\limits_{n\to \infty}ce^{-1}\sqrt{2\pi
(d-1)}\left(\frac{e^{\delta}}{n^{1-\alpha-\delta}}\right)^{d-1}\\
&=&0.
\end{eqnarray*}
\end{prf}

\begin{lem}\label{lem:d=n^a:poly}
For every constants $0<\alpha<1$ and $\epsilon>0$, whenever
$d=n^{\alpha}$,
\[
\lim_{n\to\infty}\frac{d^2+d}{D^{\epsilon}}=0.
\]
\end{lem}
\begin{prf}We have
\begin{eqnarray*}
\lim_{n\to\infty}\frac{d^2+d}{D^{\epsilon}}=
\lim_{n\to\infty}\frac{n^{2\alpha}+n^{\alpha}}{D^{\epsilon}}
&\leq&
 \lim_{n\to\infty}\frac{n^{2\alpha}+n^{\alpha}}{\left({n\choose
 d-1}D_{d-1}\right)^{\epsilon}}\\
 &=&\lim_{n\to\infty}\frac{n^{2\alpha}+n^{\alpha}}{\left(\frac{n!(d-1)!}{e(d-1)!(n-d+1)!}\right)^{\epsilon}}\\
 &\leq&\lim_{n\to\infty}\frac{n^{2\alpha}+n^{\alpha}}{\left(\frac{(n-d+2)^{d-1}}{e}\right)^{\epsilon}}\\
 &=&\lim_{n\to\infty}\frac{(n^{2\alpha}+n^{\alpha})e^{\epsilon}}{\left(n-n^{\alpha}+2\right)^{(n^{\alpha}-1)\epsilon}}\\
 &=&0.
\end{eqnarray*}
\end{prf}

We are now ready to complete the proof of
Lemma~\ref{lem:relation_permutation_Gilbert_2}. It follows from
Lemma~\ref{lem:relation_binary_Gilbert:d=n^a} that there is a
positive constant $\varepsilon$ satisfying $T'\leq
D'^{2-\varepsilon}$. This combing with
Lemma~\ref{lem:relation:T:T' and D'}, we obtain
\begin{eqnarray}
T&\leq& (T'+D')D^2_{d-1}\nonumber\\
&\leq& (D'^{2-\varepsilon}+D')D^2_{d-1}\nonumber\\
&=&(D'D_{d-1})^{2-\varepsilon}D_{d-1}^{\varepsilon}+(D'D_{d-1})D_{d-1}\label{eq:lem:relation_permutation_Gilbert_23}
\end{eqnarray}
It follows  from
Lemma~\ref{lem:forprf:lem:relation_permutation_Gilbert_2} that for
any constant $0<\delta<1-\alpha$,  there exists a positive
constant $N$, for $n\geq N$ satisfying
\begin{equation}\label{eq:lem:relation_permutation_Gilbert_20}
D_{d-1}<D^{1-\delta},
\end{equation}
and follows from the definitions of $D,D',D_{d-1}$ that
\begin{equation}\label{eq:lem:relation_permutation_Gilbert_21}
D'D_{d-1}\leq d{n\choose d-1}D_{d-1}\leq dD.
\end{equation}
By applications of (\ref{eq:lem:relation_permutation_Gilbert_20})
and (\ref{eq:lem:relation_permutation_Gilbert_21}) for
(\ref{eq:lem:relation_permutation_Gilbert_23}), we have
\begin{eqnarray}
T&\leq&(dD)^{2-\varepsilon}D^{\varepsilon(1-\delta)}+dDD^{1-\delta}\nonumber\\
&\leq&(d^2+d)D^{2-\varepsilon\delta},
\label{eq:lem:relation_permutation_Gilbert_26}
 \end{eqnarray}
By Lemma~\ref{lem:d=n^a:poly} there exists a positive constant
$M$, for $n\geq M$ satisfying $d^2+d\leq D^{\varepsilon\delta/2}$.
This in conjunction with
(\ref{eq:lem:relation_permutation_Gilbert_26}) follows that for
$n\geq \max(N,M)$, $T\leq D^{2-\varepsilon\delta/2}$. Since
$T<D^2$ always holds, there exists a positive constant
$\varepsilon'$ for $0< n<\max(N,M)$ satisfying $T\leq
D^{2-\varepsilon'}$. Therefore taking
$\epsilon=\min(\varepsilon\delta/2,\varepsilon')$, for all $n$,
$T\leq D^{2-\epsilon}$.
\end{prf}

\emph{Proof of Theorem~\ref{thm:graph:d:small}} We are now ready
to complete the proof of Theorem~\ref{thm:graph:d:small}. Let
$\alpha$ be a constant satisfying  $0<\alpha <1$. Then by the
definitions of $D$ and $T$, Theorem~\ref{thm:general-graph} and
Lemma~\ref{lem:relation_permutation_Gilbert_2},  for case
$d=n^{\alpha}$ there exists an positive constant $\epsilon$ such
that
\begin{equation*}
\begin{array}{lcl}
\alpha(G_{P})&\geq&
\frac{n!}{10D}\left(\log_2D-1/2\log_2\left(\frac{D^{2-\epsilon}}{3}\right)\right)\\
&\geq&\frac{\min(\epsilon,1)}{20}\cdot \frac{n!}{D}(\log_2 D+\log_23)\\
&\geq&\frac{\min(\epsilon,1)}{20}\cdot \frac{n!}{V(n,d-1)}\log_2
V(n,d-1).
\end{array}
\end{equation*}
Then we complete the proof.

\hfill QED.

\section{Improved the Gilbert-Varshamov Bound by Considering Covered Balls intersections}
A directly approach to improve the Gilbert-Varshamov bound is to
consider the intersections of the covered balls of the codewords.
By this approach, two bounds depended on other quantities are
given in this section.

\begin{thm}\label{thm:P(n,d)-P[n,d-1]}
\begin{equation}
P(n,d)\geq \frac{2\cdot n!}{V(n,d-1)+P[n,d-1]}
\end{equation}
\end{thm}
\begin{prf}Let $C$ be an $(n,P(n,d),d)$ PA.
Let $\mathbf{c}\in C$. Suppose $\mathbf{a}$ and $\mathbf{b}$ are
two distinct permutations covered by $\mathbf{c}$ only. Then it
must have $d(\mathbf{a},\mathbf{b})<d$, otherwise $C\cup
\{\mathbf{a},\mathbf{b}\}/\{\mathbf{c}\}$ is an $(n,d)$ PA of size
$P(n,d)+1$, which is a contradiction. This implies there are at
most $P[n,d-1]$ permutations covered by $\mathbf{c}$ only. Then
there are at least $n!-P(n,d)P[n,d-1]$ permutations in $S_n$
covered by at least 2 codewords. So we have
\begin{eqnarray*}
P(n,d)V(n,d-1)&=& \sum_{\mathbf{c}\in
C}|\mathbf{B}(\mathbf{x})|\\
&\geq&n!+|\{\mathbf{a}\in S_n: \mathbf{a}\mbox{ is covered by at
least two
codewords.} \}|\\
&\geq&n!+n!-P(n,d)P[n,d-1],
\end{eqnarray*}
which implies the claim of the theorem.
\end{prf}

Clearly, $P[n,d-1]\leq V(n,d-1)$, then the bound in
Theorem~\ref{thm:P(n,d)-P[n,d-1]} is an improvement over the
Gilbert-Varshamov bound for PA. While determining the exact values
of $P[n,d-1]$ seems difficult, for $n$ being small values, the
upper bounds on $P[n,d-1]$ can be obtained by linear programming
\cite{Tarnanen99}, for general cases, bounds on $P[n,d-1]$ are
given below.
\begin{prop}
For all $d\leq n$,
\[
P[n,d-1]\geq\max\{(d-1)!, V(n,\lfloor (d-1)/2\rfloor)\},
\]
moreover for $d$ being even,
\[
P[n,d-1]\geq V(n,d/2-1)+{n-1\choose d/2-1}D_{d/2}.
\]
For all $d\leq n$,
\[
P[n,d-1]\leq \max\left\{L_{i,0}+\ldots+L_{i,i}: i=\lfloor
(d-1)/2\rfloor,\ldots,d-1\right\},
\]
where $L_{i,j}$ is defined in Lemma~\ref{lem:bound:L_i,j}.
\end{prop}
\begin{prf}
Clearly, the set of permutations with supports be subsets of
$\{0,1,\ldots,d-2\}$ has pairwise distances less than $d$. This
implies $P[n,d-1]\geq |S_{d-1}|=(d-1)!$. And the set
\[
A=\{\mathbf{x}\in S_n: wt(\mathbf{x})\leq \lfloor (d-1)/2\rfloor\}
\]
has pairwise distances less than $d$ also. This lead to
$P[n,d-1]\geq |A|=V(n,\lfloor (d-1)/2\rfloor)$. For case $d$ being
even, the set
\[
B=\{\mathbf{x}\in S_n:wt(\mathbf{x})=d/2,\mathbf{x}(0)\neq 0\}
\]
has pairwise distances less than $d$, moreover the distance from
 any permutation in $A$ to any permutation in $B$ is less than $d$. Hence
For case $d$ being even, $P[n,d-1]\geq
|A|+|B|=V(n,d/2-1)+{n-1\choose d/2-1}D_{d/2}$.

 Suppose $C$ is a subset of $S_n$ with size of $P[n,d-1]$ and pairwise distances less than $d$.
Without loss of generality, we assume that $\mathbf{1}$ is an
element of $C$. If the maximum weight of permutations in $C$ is
$i$, then $|C|\leq L_{i,0}+\ldots+L_{i,i}$ by
Lemma~\ref{lem:bound:L_i,j}. If $i=\lfloor (d-1)/2\rfloor$ then
$C$ includes all the permutations with weights not more than
$\lfloor (d-1)/2\rfloor$. Therefore we obtain the upper bound on
$P[n,d-1]$ presented in the proposition.
\end{prf}

\emph{Remark:} Another connection between $P(n,d)$ and $P[n,d-1]$
shown in ~\cite[Theorem 3]{Tarnanen99} is that
\[
P(n,d)P[n,d-1]\leq n!.
\]

%%%%%%%%%%%%%%%%%%%%%%%%%%%%%%%%%%%%%%%%%%%%%%%%%%%%%%%%%%%%%%%%%%%%
\begin{thm}
Let $C'$ be an $(n,M,d)$ PA, then
\[
P(n,d)\geq \frac{n!M}{|\mathbf{B}(C')|}.
\]
\end{thm}
\begin{prf}
Suppose $C$ is an $(n,P(n,d),d)$ PA. Then for any $\mathbf{x}\in
S_n$, $(\mathbf{x}C/(\mathbf{x}C\cap \mathbf{B}(C')))\cup C'$ is
an $(n,d)$ PA with size
$|\mathbf{x}C|-|\mathbf{x}C\cap\mathbf{B}(C')|+|C'|$, where
$\mathbf{x}C=\{\mathbf{x}\mathbf{c}:\mathbf{c}\in C\}$. Clearly,
$|\mathbf{x}C|-|\mathbf{x}C\cap\mathbf{B}(C')|+|C'|\leq P(n,d)$.
 This in
conjunction with $|\mathbf{x}C|=|C|=P(n,d)$ and $|C'|=M$ yields
$P(n,d)-|\mathbf{x}C\cap\mathbf{B}(C')|+M\leq P(n,d)$, i.e.
$|\mathbf{x}C\cap\mathbf{B}(C')|\geq M$. Then $\sum_{\mathbf{x}\in
S_n}|\mathbf{x}C\cap\mathbf{B}(C')|\geq Mn!$. On the other hand,
we have
\begin{eqnarray*}
\sum_{\mathbf{x}\in S_n}|\mathbf{x}C\cap\mathbf{B}(C')|
&=&\sum_{\mathbf{b}\in \mathbf{B}(C')}\sum_{\mathbf{c}\in C}|\{\mathbf{x}\in S_n:\mathbf{x}\mathbf{c}=\mathbf{b}\}|\\
&=&\sum_{\mathbf{b}\in \mathbf{B}(C')}\sum_{\mathbf{c}\in C}1\\
&=&|\mathbf{B}(C')|P(n,d)
\end{eqnarray*}
Therefore $Mn!\leq |\mathbf{B}(C')|P(n,d)$, in other words
$P(n,d)\geq \frac{n!M}{|\mathbf{B}(C')|}$.
\end{prf}

Since
\[
\frac{\frac{n!M}{|\mathbf{B}(C')|}}{\frac{n!}{V(n,d-1)}}=\frac{M\cdot
V(n,d-1)}{|\mathbf{B}(C')|}=\frac{\sum_{\mathbf{c}\in
C'}|\mathbf{B}(\mathbf{c})|}{|\cup_{\mathbf{c}\in
C'}\mathbf{B}(\mathbf{c})|},
\]
we can expect to improve the Gilbert-Varshamov bound on $P(n,d)$
by constructing an $(n,d)$ PA with relative small size of covered
ball. For instance, in~\cite[Section 1, p.54]{wensong04}, it is
suggested to choose $d$ permutations with pairwise distance
exactly $d$. But evaluation of $|B(C')|$ seems difficult.

\section{Lower Bounds for Certain Cases}

In this section, some new lower bounds for certain values of $n$
and $d$ are given. These new bounds follow from two inequalities
in $P(n,d)$ which are derived by two constructions as follows,
respectively.

\begin{lem}\label{lem:d:d-3}
Suppose $n\geq d> 3$. Let $\Phi=\{\phi_i\}_{i=1}^M$ be an
$(n,M,d)$ PA, and let $\psi_i:\mathbf{Z}_{n-1}\mapsto
\mathbf{Z}_{n-1}$ be defined as follows
\[
\psi_i(x)=\left\{\begin{array}{cc} \phi_i(x),&\mbox{for
}\phi_i(x)\neq n-1\\
\phi_i(n-1),&\mbox{for }\phi_i(x)=n-1.
\end{array}\right.
\]
Then $\Psi=\{\psi_i\}_{i=1}^M$ forms an $(n-1,M,d-3)$ PA.
\end{lem}

\begin{prf}
Obviously, each $\psi_i\in\Psi$ is a permutation over
$\mathbf{Z}_{n-1}$. For any  $1\leq i,j\leq M,i\neq j$, if
$\phi_i(x)\neq n-1,\phi_j(x)\neq n-1$, then $\psi_i(x)\neq
\psi_j(x)$ if and only if $\phi_i(x)\neq\phi_j(x)$, thus we have
\[
d(\psi_i,\psi_j)\geq |\{x:x\in Z_{n-1},\phi_i(x)\neq
n-1,\phi_j(x)\neq n-1,\phi_i(x)\neq \phi_i(x)\}|\geq
d(\phi_i,\phi_j)-3\geq d-3,
\]
which implies the statement.
\end{prf}

\begin{lem}\label{lem:d:d-2}
Suppose $n\geq d> 2$. Let $\Phi$ be an $(n,M,d)$ PA, and let
$\Phi_i=\{\phi\in \Phi: \phi(i)=n-1\},i=0,\ldots,n-1$. Suppose for
$s\neq t$ and for any $k\neq s,t$, $|\Phi_s|\geq |\Phi_t|\geq
|\Phi_k|$, and $\Phi_s=\{\phi_{i}^s\}_{i=1}^{M_1},
\Phi_t=\{\phi_{j}^t\}_{j=1}^{M_2}$. Let
$\psi^s_i:\mathbf{Z}_n/\{s\}\mapsto \mathbf{Z}_n/\{s\}$ and
$\psi^t_j:\mathbf{Z}_n/\{s\}\mapsto \mathbf{Z}_n/\{s\}$ be defined
respectively as follows
\[
\psi^s_i(x)=\phi^s_i(x),\mbox{for }x\in \mathbf{Z}_n/\{s\},
\]
\[
\psi^t_j(x)=\left\{\begin{array}{cl} \phi^t_j(x),&\mbox{for
}x\in \mathbf{Z}_n/\{s,t\}\\
\phi^t_j(s),&\mbox{for }x=t.
\end{array}\right.
\]

Then $\Psi=\{\psi^s_i\}_{i=1}^{M_1}\cup\{\psi^t_j\}_{j=1}^{M_2}$
is an $(n-1,d-2)$ PA over $\mathbf{Z}_n/\{s\}$ of size
$M_1+M_2\geq \frac{2M}{n}$.
\end{lem}

\begin{prf}
Obviously, each $\psi^s_i\in\Psi$ and each $\psi^t_j\in\Psi$ are
permutations over $\mathbf{Z}_{n}/\{s\}$. Moreover, for any
permutations $\psi_i^s,\psi_j^t\in \Psi$ and any $x\in
Z_n/\{s,t\}$, $\psi_i^s(x)=\phi_i^s(x),\psi_j^t(x)=\phi_j^t(x)$.
So for any distinct permutations
$\psi_i^s,\psi_j^s,\psi_{i'}^t,\psi_{j'}^t\in \Psi$,
$d(\psi_i^s,\psi_j^s)\geq d(\phi_i^s,\phi_j^s)-2\geq d-2$,
$d(\psi_{i'}^t,\psi_{j'}^t)\geq d(\phi_{i'}^t,\phi_{j'}^t)-2\geq
d-2$ and $d(\psi_i^s,\psi_{i'}^t)\geq
d(\phi_i^s,\phi_{i'}^t)-2\geq d-2$. Hence the lemma immediately
follows.
\end{prf}

From Lemma~\ref{lem:d:d-3} and Lemma~\ref{lem:d:d-2} we have the
following theorem immediately.
\begin{thm}\label{thm:bound:d-3}
For $n\geq d> 3$
\begin{equation}\label{bound:relate}
P(n-1,d-3)\geq P(n,d).
\end{equation}

For $n\geq d> 2$
\[
P(n-1,d-2)\geq \frac{2}{n}P(n,d).
\]
\end{thm}

\begin{cor}\label{cor:special:newlowerbound}
Let $q$ be the power of prime number. Then
\begin{eqnarray*}
P(q,q-4)&\geq&(q+1)q(q-1)\\
P(q,q-3)&\geq&2q(q-1)\\
P(q-1,q-4)&\geq& (q+1)(q-1)\\
P(q-1,q-6)&\geq& 2(q+1)(q-1).
\end{eqnarray*}
Additionally, $P(11,5)\geq 95040$ and $P(11,6)\geq 15840$.
\end{cor}
\begin{prf}
If $q$ is a prime-power, then it follows from
Theorem~\ref{thm:bound:d-3} and Proposition~\ref{prop:PA_sharply}
that
\begin{equation}\label{prf:cor:low_bound_sepcial}
\begin{array}{lcl}
P(q,q-4)&\geq&P(q+1,q-1)\\
&=&(q+1)q(q-1)\\
\end{array}
\end{equation}
and $P(q,q-3)\geq
\frac{2}{q+1}P(q+1,q-1)=\frac{2(q+1)q(q-1)}{q+1}=2q(q-1)$.
Moreover $P(11,5)\geq P(12,8)=95040$,
$P(11,6)\geq\frac{2}{12}P(12,8) =\frac{2\cdot 95040}{12}=15840$.
(\ref{eq:n,n-1:prop:elementay:conse:P(n,d)}) in conjunction with
(\ref{prf:cor:low_bound_sepcial}), yields
\begin{equation*}
\begin{array}{lcl}
P(q-1,q-4)&\geq&\frac{1}{q}P(q,q-4)\\
&\geq&(q+1)(q-1).
\end{array}
\end{equation*}
Additionally, Theorem~\ref{thm:bound:d-3} in conjunction with
(\ref{prf:cor:low_bound_sepcial}), yields
\begin{equation*}
\begin{array}{lcl}
P(q-1,q-6)&\geq&\frac{2}{q}P(q,q-4)\\
&\geq&2(q+1)(q-1).
\end{array}
\end{equation*}
\end{prf}

In general, for certain cases, the lower bounds given by
Corollary~\ref{cor:special:newlowerbound} are more tighter than
the previous bounds, and they are compared in
Table~\ref{table:compare:lowerbound:1} and
~\ref{table:compare:lowerbound:2}, where the function $\theta(x)$
in Table~\ref{table:compare:lowerbound:2} is defined by
(\ref{eq:theta_function}). Moreover, The new bounds $P(11,5)\geq
95040$ and $P(11,6)\geq 15840$ are also tighter than the previous
bound $P(11,5)\geq 60940$ and $P(11,6)\geq 9790$~\cite[Table 5,
p.63]{wensong04} respectively.

\begin{table}
  \centering
  \caption{Comparison of lower bounds on $P(q,q-3)$ and $P(q,q-4)$}\label{table:compare:lowerbound:1}
  \begin{tabular}{|l|l|l|l|l|}
    \hline
    \multicolumn{1}{|c|}{}&\multicolumn{1}{|c|}{}&\multicolumn{3}{|c|}{Lower bound on $P(q,d)$}\\
    \cline{3-5}$q$ is a prime power & $d$ & Corollary 2& Proposition 4& Proposition 5\\
    \hline
    $q\equiv 1\mod 6,q\neq 7$&$q-3$&$2q(q-1)$&$q(q-1)$&$q$\\
    $q\equiv 1\mod 6$ and $q\equiv 0\mod 5$  & $q-4$ &$(q+1)q(q-1)$ & $q(q-1)$ & $\frac{1}{2}q^3+\frac{1}{4}q^2+\frac{5}{4}q$\\
    $q\equiv 1\mod 6$ and $q\equiv 1\mod 5$  & $q-4$ &$(q+1)q(q-1)$ & $q(q-1)$ & $q$\\
    $q\equiv 1\mod 6$ and $q\equiv -1\mod 5$  & $q-4$ &$(q+1)q(q-1)$ & $q(q-1)$ & $q^2+q$\\
    \hline
  \end{tabular}
\end{table}

\begin{table}
  \centering
  \caption{Comparison of lower bounds on $P(q-1,q-4)$}\label{table:compare:lowerbound:2}
  \begin{tabular}{|l|l|l|l|l|}
    \hline
    \multicolumn{1}{|c|}{}&\multicolumn{4}{|c|}{Lower bound on $P(q-1,q-4)$}\\
    \cline{2-5}$q$ is a prime power& Corollary 2& Proposition 4 & Proposition 5&Proposition~\ref{prop:n,n-1}\\
    &&and (\ref{eq:n,n-1:prop:elementay:conse:P(n,d)})&and (\ref{eq:n,n-1:prop:elementay:conse:P(n,d)})&and (\ref{eq:n,d,d+1})\\
    \hline
    $q\equiv 1\mod 6$ and $q\equiv 0\mod 5$  &$(q+1)(q-1)$ & $q-1$ & $\frac{1}{2}q^2+\frac{1}{4}q+\frac{5}{4}$&$(q-1)(\theta(q-1)-1)$\\
    $q\equiv 1\mod 6$ and $q\equiv 1\mod 5$  &$(q+1)(q-1)$ & $q-1$ & $1$&$(q-1)(\theta(q-1)-1)$\\
    $q\equiv 1\mod 6$ and $q\equiv -1\mod 5$ &$(q+1)(q-1)$ & $q-1$ &$q+1$&$(q-1)(\theta(q-1)-1)$\\
    \hline
  \end{tabular}
\end{table}

 \hfill mds

\hfill November 18, 2002

\bibliographystyle{IEEEtran}
\bibliography{IEEEabrv,bib}
\end{document}